\documentstyle[12pt]{article}
\def\be{\begin{equation}}
\def\ee{\end{equation}}
\def\bea{\begin{eqnarray}}
\def\eea{\end{eqnarray}}

\begin{document}

\begin{center}
{\Large{\bf Noncommutativity Parameter As a Field on the String Worldsheet}}

\vskip .5cm
{\large Davoud Kamani}
\vskip .1cm
 {\it Institute for Studies in Theoretical Physics and Mathematics (IPM)
\\  P.O.Box: 19395-5531, Tehran, Iran}\\
{\sl e-mail: kamani@theory.ipm.ac.ir}
\\
\end{center}

\begin{abstract}

We consider the noncommutativity parameter of the space-time 
as a bosonic worldsheet field.
By finding a fermionic super-partner for it, we can find star products
between the boson-boson, boson-fermion and fermion-fermion fields
of superstring worldsheet
and also between superfields of the worldsheet superspace. We find a 
two dimensional action for the noncommutativity 
parameter and its fermionic partner. We discuss the symmetries of this action.

\end{abstract}
\newpage
\section{Introduction}
The study of open strings in the presence of background fields enables us to
explain the noncommutativity on the D-brane worldvolume \cite{1,2}. In the
most of these attempts the noncommutativity parameter is constant i.e., it is
independent of the spacetime coordinates. It is given in terms of a 
constant background NS$\otimes$NS $B$-field and a
constant closed string metric $g_{\mu \nu}$ \cite{1}. The noncommutativity
parameter also can be non-constant \cite{3,4,5}. One can achieve to this, by
introducing a nontrivial $B$-field \cite{4} or more general, a curved 
background metric
$g_{\mu \nu}$ with nontrivial $B$-field \cite{5}. 

The fact that 
the noncommutativity parameter is not constant implies that, indirectly 
this parameter depends on the coordinates of the string worldsheet.
We consider the noncommutativity parameter $\Theta^{\mu \nu}$ as a bosonic
field of the string worldsheet. Worldsheet supersymmetry
enables us to find a fermionic super-partner for it. We call this 
fermionic field $\omega^{\mu \nu}$. 
We study supersymmetry of the $\Theta \omega$-system. That
is, we find the supersymmetry transformations of the fields 
$\Theta^{\mu \nu}$ and $\omega^{\mu \nu}$. We show that the algebra
of these transformations is closed. 

Imposing the worldsheet
supersymmetry to the noncommutativity relation of the spacetime, creates 
star products between the boson-boson, boson-fermion and fermion-fermion
fields. By these products we can find the noncommutativity of the superfields
associated to the worldsheet superspace. Therefore, this noncommutativity 
is compatible with supersymmetry. 
Furthermore, we shall show that star product
of two superfields is a superfield. In other words, noncommutativity
parameter of the superfield coordinates is a superfield.

Some transformation properties of the components of the supersymmetrized
local noncommutativity parameter will be discussed.
By quantizing the system, the 
algebra of the supercurrent is studied.
In addition to the supersymmetry, the action of the 
$\Theta \omega$-system also is symmetric
under the reparametrization of the worldsheet coordinates, the Weyl scaling
of the worldsheet metric, the transformations that are induced by 
Poincar\'e transformations and 
another linear transformation of $\Theta^{\mu \nu}$ and $\omega^{\mu \nu}$. 
We study the two later symmetries in detail.

This paper is organized as follows. In section 2,
supersymmetry of the noncommutativity parameter and its fermionic partner is
discussed. In section 3, we find star products
between the worldsheet bosons and fermions and also between superfields of
worldsheet superspace. In section 4, the action of these new fields,
the quantization of these fields, the supercurrent and its algebra will 
be obtained. In section 5, we study the induced Poincar\'e symmetry and 
another symmetry of this system.
\section{Supersymmetry transformations of $\Theta^{\mu \nu}$ and 
$\omega^{\mu \nu} $} 

We know that for a noncommutative space with the coordinates $\{X^\mu\}$
and the noncommutativity parameter $\Theta^{\mu \nu}$ there is
\bea
X^\mu * X^\nu = X^\mu X^\nu + \frac{i}{2} \Theta^{\mu \nu}\;,
\eea
which leads to the commutator
\bea
[X^\mu\;,\; X^\nu ]_* = i \Theta^{\mu \nu}\;.
\eea
Note that in string theory, the equation (1) is definition of the star 
product between two worldsheet bosons $X^\mu$ and $X^\nu$. 
It is not definition of the parameter $\Theta^{\mu \nu}$.
Later we shall find an appropriate action for $\Theta^{\mu \nu}$.

Now consider superstring theory with worldsheet supersymmetry. 
It has the action  
\bea
S=-\frac{1}{4\pi \alpha'} \int d^2 \sigma
( \partial_a X^\mu \partial^a X_\mu - i {\bar \psi}^\mu \rho^a \partial_a
\psi_\mu )\;,
\eea
where 
$\psi^\mu = \left( \begin{array}{c}
\psi^\mu_-  \\ 
\psi^\mu_+
\end{array} \right)$ 
is a Majorana spinor of worldsheet. The equations of motion
extracted from this action are
\bea
\partial_+ \psi^\mu_- = \partial_- \psi^\mu_+ = \partial_+ \partial_- 
X^\mu=0\;,
\eea
where $\partial_{\pm} = \frac{1}{2} (\partial_\tau \pm \partial_\sigma)$.
Under the worldsheet supersymmetry transformations, this action is invariant.
These transformations are
\bea
&~& \delta X^\mu = {\bar \epsilon}\psi^\mu\;,
\nonumber\\
&~& \delta \psi ^\mu = -i \rho^a \partial_a X^\mu \epsilon \;,
\eea
where $\epsilon$ is a constant anti-commuting infinitesimal Majorana spinor. 
According to the equations of motion (4),  
supersymmetry means that $\partial_{\pm} X^\mu$ appears like 
$\psi^\mu _{\pm}$ and vice-versa.

From the equation (1) and the supersymmetry transformations (5), we have
\bea
\frac{i}{2} \delta \Theta^{\mu \nu} = {\bar \epsilon} (
\psi^\mu * X^\nu + X^\mu * \psi^\nu
-\psi^\mu  X^\nu - X^\mu  \psi^\nu)\;.
\eea
This transformation can be written as
\bea
\delta \Theta^{\mu \nu} = {\bar \epsilon} \omega^{\mu \nu}\;,
\eea
where $\omega^{\mu \nu}$ is defined as
\bea
\frac{i}{2} \omega^{\mu \nu}
=\psi^\mu * X^\nu + X^\mu * \psi^\nu -\psi^\mu X^\nu - X^\mu  \psi^\nu \;,  
\eea
or in the commutator form, it is
\bea
i\omega^{\mu \nu}
=[ \psi^\mu\;,\; X^\nu ]_* - [\psi^\nu\;,\;X^\mu]_*\;.  
\eea
Also the equations (5) and (8) give the supersymmetry transformation of 
$\omega^{\mu \nu}$ i.e.,
\bea
\frac{i}{2} \delta \omega^{\mu \nu} = -i \rho^a \epsilon ( 
\partial_a X^\mu * X^\nu + X^\mu * \partial_a X^\nu
-\partial_a X^\mu  X^\nu - X^\mu  \partial_a X^\nu )\;.
\eea
By using the equation (1) this transformation takes the simple form
\bea
\delta \omega^{\mu \nu} = -i \rho^a \partial_a \Theta^{\mu \nu}\epsilon\;.
\eea
Note that the transformations of the equations (2) and (9) also lead to the
results (7) and (11), respectively.
Now we have a new Majorana spinor
$\omega^{\mu \nu} = \left( \begin{array}{c}
\omega^{\mu \nu}_-  \\ 
\omega^{\mu \nu}_+
\end{array} \right)$ 
which is antisymmetric under the exchange of the indices $\mu$ and $\nu$. 

The transformations (7) and (11) form a closed algebra, that is for two 
successive transformations $\delta_\epsilon$ and 
$\delta_{\epsilon'}$ we obtain
\bea
&~&[\delta_\epsilon\;,\;\delta_{\epsilon'}]\Theta^{\mu \nu}
=2i {\bar \epsilon} \rho^a \epsilon' \partial_a  \Theta^{\mu \nu}\;,
\nonumber\\
&~&[\delta_\epsilon\;,\;\delta_{\epsilon'}]\omega^{\mu \nu} 
= 2i {\bar \epsilon} \rho^a \epsilon' \partial_a  \omega^{\mu \nu}\;.
\eea
Note that closeness of algebra means that the commutator of two 
supersymmetry transformations gives a spatial translation. 
This can be seen from the above equations.

\section{Star product between various fields}

Now we find star product between the worldsheet fields and between
superfields.
Making derivative of the equation (1) with respect to the light-cone
coordinates $\sigma^{\pm}$, leads to the equation
\bea
\frac{i}{2} \partial_{\pm}\Theta^{\mu \nu}=
\partial_{\pm} X^\mu * X^\nu + X^\mu * \partial_{\pm} X^\nu 
-\partial_{\pm} X^\mu  X^\nu - X^\mu  \partial_{\pm} X^\nu\;. 
\eea
According to the supersymmetry this equation gives
\bea
\frac{i}{2} \Omega^{\mu \nu} 
=\psi^\mu* X^\nu + X^\mu * \psi^\nu - \psi^\mu X^\nu - X^\mu  \psi^\nu \;, 
\eea
where we introduced the spinor $\Omega^{\mu\nu}$ as super-partner of
$\Theta^{\mu\nu}$. The equations (8) and (14) give the equality 
$\Omega^{\mu\nu}=\omega^{\mu\nu}$. Therefore, the equations (7), (8)
and (11) imply $\Theta^{\mu\nu}$ and $\omega^{\mu\nu}$ are super-partner
of each other. Again note that the equation (8) is
definition of star product between the worldsheet bosons $\{ X^\mu \}$ and
the worldsheet fermions $\{\psi^\mu \}$. 
It is not definition of the spinor $\omega^{\mu \nu}$.
We shall give an appropriate action for $\omega^{\mu \nu}$. 

From the equation (13) we obtain
\bea
\partial_+ X^\mu * \partial_- X^\nu + \partial_- X^\mu * \partial_+ X^\nu =
\partial_+ X^\mu  \partial_- X^\nu + \partial_- X^\mu  \partial_+ X^\nu 
+\frac{i}{2} \partial_+ \partial_- \Theta^{\mu \nu}\;, 
\eea
Using the supersymmetry, leads to the equation
\bea
\psi^\mu_+ * \psi^\nu_- + \psi^\mu_- * \psi^\nu_+
=\psi^\mu_+  \psi^\nu_- + \psi^\mu_-  \psi^\nu_+ 
+\frac{i}{2} \partial_+ \partial_- \Theta^{\mu \nu}\;.
\eea
This shows the star product between the components of the spinor fields of
the string worldsheet. This product, similar to the equation (8), naturally
is defined by two terms that contain star product. The equation (16) also
can be obtained from the equation (8).

According to the equations (1), (8) and (16) we can find star product and
noncommutativity parameter for a space with worldsheet superfields as its 
coordinates. A superfield in general has the form
\bea
Y^\mu = X^\mu + {\bar \theta} \psi^\mu + \frac{1}{2} {\bar \theta}
\theta B^\mu\;,
\eea
where the Grassmann coordinates $\theta^1$ and $\theta^2$ form a two 
dimensional Majorana spinor 
$\theta = \left( \begin{array}{c}
\theta^1 \\ 
\theta^2
\end{array} \right)$. 
Under the supersymmetry, this field transforms as
$\delta Y^\mu = {\bar \epsilon}Q(Y^\mu)$, where the generator
$Q = \frac{\partial}{\partial {\bar \theta}}
+ i\rho^a \theta \partial_a$
represents supersymmetry on the superspace. 
The commutator of two superfields with star product is
\bea
[Y^\mu\;,\; Y^\nu ]_* = i\Phi^{\mu \nu} \equiv 
i(\Theta^{\mu \nu} + {\bar \theta}\omega^{\mu \nu}
+ \frac{1}{2}{\bar \theta} \theta \Gamma^{\mu \nu}) \;,
\eea
where the matrix $\Gamma^{\mu \nu}$ is 
\bea
\Gamma^{\mu \nu} = i({\bar \psi}^\mu * \psi^\nu
-{\bar \psi}^\nu * \psi^\mu)\;.
\eea
The antisymmetric tensor $\Phi^{\mu\nu}$ shows the noncommutativity 
parameter of the space $\{Y^\mu\}$. 
This parameter depends on the superspace coordinates $\sigma, \tau, \theta^1$
and $\theta^2$. The various components of $\Phi^{\mu \nu}$ i.e., 
$\Theta^{\mu \nu}$, $\omega^{\mu \nu}$ and $\Gamma^{\mu \nu}$ show the
noncommutativity of $X-X$, $X-\psi$ and $\psi-\psi$ subspaces.

We know that usual product of two superfields $Y^\mu$ and $Y^\nu$ is a
superfield i.e., $\delta (Y^\mu Y^\nu)={\bar \epsilon}Q (Y^\mu Y^\nu)$.
This also holds by star product i.e., $Y^\mu * Y^\nu$ is a superfield.
Now we show this. The star product is
\bea
Y^\mu * Y^\nu = Y^\mu Y^\nu + \frac{i}{2} \Lambda^{\mu\nu}\;,
\eea
where $\Lambda^{\mu\nu}$ is
\bea
&~&\Lambda^{\mu\nu}= \Theta^{\mu\nu}+ {\bar \theta} \omega^{\mu\nu}
+\frac{1}{2}{\bar \theta}\theta \lambda^{\mu\nu}\;,
\nonumber\\
&~&\lambda^{\mu\nu}=2i({\bar \psi^\mu}*\psi^\nu-{\bar \psi^\mu}\psi^\nu)\;.
\eea
Since $Y^\mu Y^\nu$ is a superfield, it is sufficient to show that
$\Lambda^{\mu\nu}$ also is a superfield. Using the equations (1) and (8)
for $\Theta^{\mu\nu}$ and $\omega^{\mu\nu}$ and also applying the geometrical
transformations $\delta \theta = \epsilon$ and 
$\delta \sigma^a = i{\bar \epsilon} \rho^a \theta$, lead to the equation
\bea
\delta \Lambda^{\mu\nu} ={\bar \epsilon}Q(\Lambda^{\mu\nu})\;,
\eea
that is, $\Lambda^{\mu\nu}$ is a superfield and therefore,
$Y^\mu * Y^\nu$ is superfield, 
\bea
\delta (Y^\mu * Y^\nu) ={\bar \epsilon}Q(Y^\mu * Y^\nu)\;.
\eea
In the same way, it is easy to show that the noncommutativity parameter 
$\Phi^{\mu\nu}$ also is superfield.

The superfield $\Lambda^{\mu\nu}$ gives the following supersymmetry 
transformations
\bea
&~& \delta \Theta^{\mu \nu} = {\bar \epsilon}\omega^{\mu\nu}\;,
\nonumber\\
&~& \delta \omega^{\mu \nu} = -i \rho^a \epsilon \partial_a \Theta^{\mu\nu} 
+ \epsilon \lambda^{\mu\nu} \;,
\nonumber\\
&~&\delta \lambda^{\mu\nu} = -i{\bar \epsilon} \rho^a \partial_a 
\omega^{\mu\nu}\;.
\eea
For compatibility with the equations (7) and (11), $\lambda^{\mu\nu}$ and
$\delta \lambda^{\mu\nu}$ should vanish i.e.,
\bea
{\bar \psi}^\mu * \psi^\nu = {\bar \psi}^\mu \psi^\nu\;,
\eea
\bea
\rho^a \partial_a \omega^{\mu\nu}=0\;.
\eea
Since $\Theta^{\mu\nu}$ is super-partner of $\omega^{\mu\nu}$, the
equation (26) implies
\bea
\partial^a \partial_a \Theta^{\mu\nu} =0\;.
\eea
The equations (16), (25) and (27) tell us that star product between
components of the worldsheet fermions $\{\psi^\mu \}$ is usual product
that is, $\psi^\mu_A * \psi^\nu_B = \psi^\mu_A \psi^\nu_B$,
where $A,B \in \{ +,- \}$.
\section{Action of the fields $\Theta^{\mu \nu}$ and $\omega^{\mu \nu}$}

According to the supersymmetry transformations (7) and (11) and
the equations (26) and (27) we can write the following action for 
$\Theta^{\mu \nu}$ and $\omega^{\mu \nu}$,
\bea
S'=-\frac{1}{4\pi \alpha'} \int d^2 \sigma
( \partial_a \Theta^{\mu \nu} \partial^a \Theta_{\mu \nu} - i 
{\bar \omega}^{\mu \nu} \rho^a \partial_a
\omega_{\mu \nu} )\;.
\eea
The equations (26) and (27) are equations of 
motion that can be extracted from
this action. The fields $\Theta^{\mu \nu}(\sigma , \tau)$ and
$\omega^{\mu \nu}(\sigma , \tau)$ are massless bosons and 
fermions that live in the worldsheet of superstring.

Under the supersymmetry transformations (7) and (11) this action is 
invariant. The supercurrent associated to this symmetry is
\bea
{\cal{J}}_a = \frac{1}{2} \rho^b \rho_a \omega^{\mu \nu}\partial_b 
\Theta_{\mu \nu}\;.
\eea
This is a conserved current i.e., $\partial^a {\cal{J}}_a =0$.
The light-cone components of this current are
\bea
&~&{\cal{J}}_+ = \omega^{\mu \nu}_+ \partial_+ \Theta_{\mu \nu}\;,
\nonumber\\
&~&{\cal{J}}_- = \omega^{\mu \nu}_- \partial_- \Theta_{\mu \nu}\;.
\eea

Now we verify the quantization of the $\Theta \omega$-system. Quantization
of the fermionic degrees of freedom is achieved by imposing the canonical
anti-commutation relations. The canonical momenta conjugate to 
$\omega^{\mu \nu}_{\pm}$ are
\bea
\Pi^{\mu \nu}_{\pm}(\sigma,\tau) = - \frac{i}{4\pi \alpha'}
\omega^{\mu \nu}_{\pm}\;.
\eea
Using the equal $\tau$ anti-commutators, we obtain
\bea
\{ \omega^{\mu \nu}_A (\sigma , \tau)\;,\; \omega^{\rho \lambda}_B 
(\sigma' , \tau) \} = \frac{\pi}{2} ( \eta^{\mu \rho} \eta^{\nu \lambda}  
- \eta^{\mu \lambda} \eta^{\nu \rho}) \delta ( \sigma - \sigma') \delta_{AB}.
\eea
Both sides under the exchanges $\mu \leftrightarrow \nu$ or 
$\rho \leftrightarrow \lambda$ change their signs. Also both sides under the
exchanges $\mu \leftrightarrow \rho\;,\; \nu \leftrightarrow \lambda\;,\;
\sigma \leftrightarrow \sigma'$ and $A \leftrightarrow B$ are invariant.
The canonical momentum conjugate to $\Theta^{\mu \nu}$ is
\bea
\Pi^{\mu \nu}(\sigma , \tau) =  \frac{1}{2\pi \alpha'}
\partial_\tau \Theta^{\mu \nu}\;.
\eea
Quantization of the bosonic degrees of freedom can be 
obtained by canonical commutation relations
\bea
[\partial_{\pm} \Theta^{\mu \nu}(\sigma , \tau)\;,\;  
\partial_{\pm} \Theta^{\rho \lambda}(\sigma' , \tau) ]  
=\pm \frac{i\pi}{4} ( \eta^{\mu \rho} \eta^{\nu \lambda}  
- \eta^{\mu \lambda} \eta^{\nu \rho}) \delta' ( \sigma - \sigma')\;,
\eea
\bea
[\partial_+ \Theta^{\mu \nu}(\sigma , \tau)\;,\;  
\partial_- \Theta^{\rho \lambda}(\sigma' , \tau) ] = 0\;. 
\eea
Under the exchanges $\mu \leftrightarrow \rho\;,\; 
\nu \leftrightarrow \lambda$
and $\sigma \leftrightarrow \sigma'$ these equations are invariant.

According to the equations (32), (34) and (35) the algebra that the 
supercurrent (30) generates, is
\bea
&~&\{ {\cal{J}}_+ (\sigma)\;,\;{\cal{J}}_+ (\sigma') \} = \pi \delta (\sigma 
-\sigma') {\cal{T}}_{++} (\sigma)\;,
\nonumber\\
&~&\{ {\cal{J}}_- (\sigma)\;,\;{\cal{J}}_- (\sigma') \} = \pi \delta (\sigma 
-\sigma') {\cal{T}}_{--} (\sigma)\;,
\nonumber\\
&~&\{ {\cal{J}}_+ (\sigma)\;,\;{\cal{J}}_- (\sigma') \} = 0\;,
\eea
where ${\cal{T}}_{++}$ and ${\cal{T}}_{--}$ are the non-zero 
light-cone components of the energy momentum tensor, extracted 
from the action (28), 
\bea
&~&{\cal{T}}_{++} = \partial_+ \Theta^{\mu \nu} \partial_+ \Theta_{\mu \nu}  
+\frac{i}{2} \omega^{\mu \nu}_+ \partial_+ \omega_{+ \mu \nu}\;, 
\nonumber\\
&~&{\cal{T}}_{--} = \partial_- \Theta^{\mu \nu} \partial_- \Theta_{\mu \nu}  
+\frac{i}{2} \omega^{\mu \nu}_- \partial_- \omega_{- \mu \nu}\;. 
\eea
The algebra (36) is similar to the algebra of the supercurrent 
associated with the supersymmetry transformations (5).
\section{Other symmetries}
The superstring action (3) under the Poincar\'e transformations 
is invariant. These transformations are
\bea
&~&\delta X^\mu = a^\mu_{\;\;\;\nu} X^\nu + b^\mu\;,
\nonumber\\
&~&\delta \psi^\mu = a^\mu_{\;\;\;\nu}\psi^\nu\;,
\eea
where $a_{\mu \nu}$ is a constant antisymmetric matrix and $b^\mu$ is a
constant vector that shows translation.

Application of the transformations (38) in the equations (1) and (8) induces 
the following transformations to the fields $\Theta^{\mu \nu}$ and 
$\omega^{\mu \nu}$, 
\bea
\delta \Theta^{\mu \nu}= (a^\mu_{\;\;\;\rho} \delta^\nu_{\;\;\;\lambda}
-a^\nu_{\;\;\;\rho} \delta^\mu_{\;\;\;\lambda}) \Theta^{\rho \lambda}\;,
\eea
\bea
\delta \omega^{\mu \nu}= (a^\mu_{\;\;\;\rho} \delta^\nu_{\;\;\;\lambda}
-a^\nu_{\;\;\;\rho} \delta^\mu_{\;\;\;\lambda}) \omega^{\rho \lambda}\;.
\eea
Since the translation $b^\mu$ is independent of the spacetime coordinates 
$\{ X^\mu \}$, it does not appear to these transformations.
Note that application of the transformations (38) in the equations (2) and 
(9) also gives the results (39) and (40).

The action (28) under the transformations (39) and (40) is invariant.
This invariance leads to the current
\bea
J^{\mu \nu \rho \lambda}_a = \frac{1}{\pi} (\Theta^{\mu \nu} \partial_a
\Theta^{\rho \lambda} - \Theta^{\rho \lambda}\partial_a \Theta^{\mu \nu}
+ i {\bar \omega}^{\mu \nu} \rho_a \omega^{\rho \lambda})\;.
\eea
Under the exchange of the indices, this current satisfies the following
identities
\bea
J^{\mu \nu \rho \lambda}_a =- J^{\nu \mu \rho \lambda}_a = 
-J^{\mu \nu \lambda \rho}_a =-J^{\rho \lambda \mu \nu}_a\;.
\eea
According to the equations of motion, this is a conserved current i.e.,
\bea
\partial^a J^{\mu \nu \rho \lambda}_a = 0\;.
\eea
Since the transformation (39) only rotates $\Theta^{\mu \nu}$ but does not
translate it, there is no current for translation.

The action (28) also is symmetric under the following linear transformations
\bea
&~&\delta \Theta^{\mu \nu}= A^{\mu \nu}_{\;\;\;\;\;\rho \lambda}\Theta^{\rho 
\lambda} + b^{\mu \nu}\;,
\nonumber\\
&~&\delta \omega^{\mu \nu}= A^{\mu \nu}_{\;\;\;\;\;\rho \lambda}\omega^{\rho 
\lambda} \;,
\eea
where $b^{\mu \nu}$ and $A^{\mu \nu}_{\;\;\;\;\;\rho \lambda}$ satisfy the 
following identities
\bea
&~&A^{\mu \nu}_{\;\;\;\;\;\rho\lambda}=- A^{\nu \mu}_{\;\;\;\;\; \rho \lambda} = 
-A^{\mu \nu}_{\;\;\;\;\lambda \rho}=-A_{\rho \lambda}^{\;\;\;\; \mu \nu}\;,
\nonumber\\
&~& b^{\mu \nu}= -b^{\nu \mu}\;.
\eea
The currents associated to the transformations (44), are the current (41)
and the current
\bea
P^{\mu \nu}_a = \frac{1}{\pi} \partial_a \Theta^{\mu \nu}\;.
\eea
This current is a result of the translation of $\Theta^{\mu \nu}$.
Also it is a conserved current i.e.,
\bea
\partial^a P^{\mu \nu}_a = 0\;.
\eea

In addition to the supersymmetry, the induced Poincar\'e symmetry and the 
symmetry under 
the transformations (44), the action (28) similar to the action (3), also
is symmetric under the reparametrization of the parameters $\sigma$ and
$\tau$ and the Weyl scaling of the worldsheet metric.
\section{Conclusions}
We considered the noncommutativity parameter as a bosonic field of 
the string worldsheet. By applying the worldsheet
supersymmetry, a super-partner was associated to it. 
We found the supersymmetry transformations of these new fields. 
Therefore, in a natural way we obtained star 
products between the boson-boson,
boson-fermion and fermion-fermion fields. According to these
products the noncommutativity parameter of the superfields 
(functions on the worldsheet superspace) was obtained. 
We saw that star product of two superfields (like usual product of them)
is a superfield. Therefore, the noncommutativity parameter of these
superfields transforms as a superfield.

From the equations of motion and the supersymmetry transformations of this 
system, we obtained action of the system.
By quantizing the system, we obtained the algebra of the supercurrent.
For this action we extracted the induced Poincar\'e symmetry and the
current associated to it. This four indices current satisfies some
identities. We showed that invariance under the transformations (44),
leads to the above four indices current and a conserved current for 
the translation of the noncommutativity parameter.
Also this action under the reparametrization of the worldsheet coordinates, 
the Weyl scaling of the worldsheet metric is invariant.

\end{document}